# Algorithm for Reproducible Analysis of Semiconducting 2D Nanomaterials Based on UV/VIS Spectroscopy

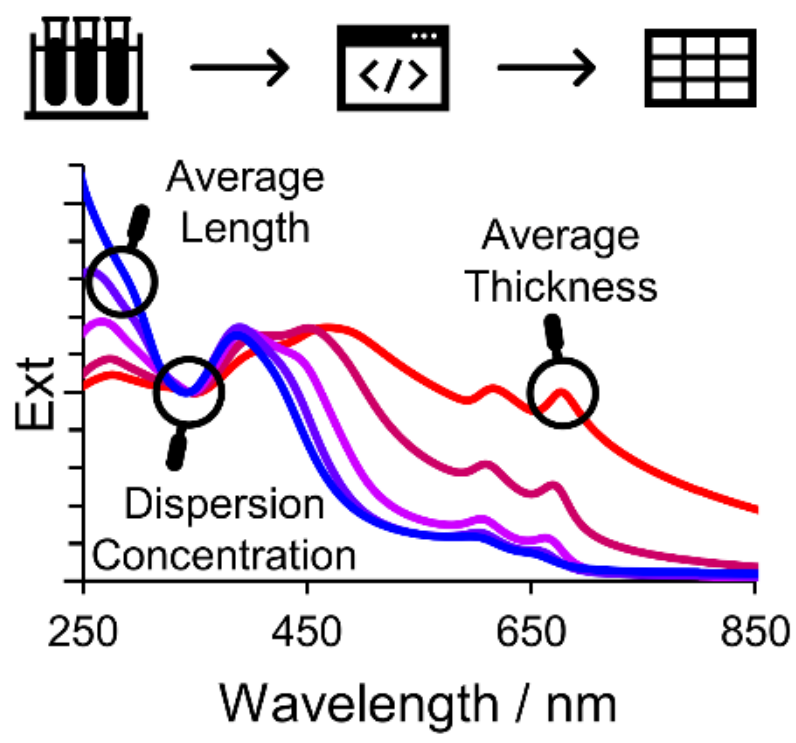

*Nico Kubetschek[1], Claudia Backes[1] and Stuart Goldie[1]**

[1] Physical Chemistry of Nanomaterials, Kassel University, Heinrich-Plett-Str. 40, 34132, Kassel, Germany

E-mail: s.goldie@uni-kassel.de

Rapid and reliable analysis of liquid dispersions of 2D materials is essential for fully harnessing their potential, allowing size and quality validation before subsequent processing or device fabrication. Existing UV/VIS extinction spectroscopy-based metrics, particularly those related to thickness, have shown promise but rely on manual data processing, which can introduce irreproducibility and user errors. To address this challenge and enable uniform analysis across laboratories, we have developed a freely available program for the reproducible analysis of nanosheet dispersions. Specifically, we apply a smoothing routine to the spectral data, take the second derivative and use integral areas to find the wavelength of exciton transitions. Our program enables rapid measurement of nanosheet concentration, length and thickness by UV/VIS spectroscopy and we have refined thickness metrics for eight common 2D nanomaterials. The program and methodology are freely available for use, and allow metrics for new materials to be implemented easily in the future.

## 1. Introduction

2D nanosheets in stable liquid dispersions have become a widely used source of nanomaterial in large quantities[1] which can be processed and used for a multitude of applications including printable electronic devices,[2,3] composite formulation,[4,5] and chemical functionalization.[6] Key to these applications are their size-dependent properties, and it is therefore important to rapidly measure their sizes in liquid dispersion.

Liquid-phase exfoliation (LPE) of 2D nanosheets was first achieved using ultrasonication in suitable solvents[7–9] and it has since been demonstrated with shear mixing[10] and microfluidization[11,12] among others. Chemical and electrochemical intercalation exfoliation methods have also seen success at producing dispersions of various 2D materials with a range of properties and potential applications. However, all produce a polydisperse mixture of nanosheet sizes and thicknesses.[13]

Microscopy, most notably transmission electron microscopy (TEM) and atomic force microscopy (AFM), is an essential tool to accurately measure the size distribution by time consuming statistical analysis, but faster methods have been developed to assess averaged properties. UV/VIS extinction and absorption spectroscopy in particular can probe the ensemble within a liquid dispersion, measuring around $10^{10}$ flakes per cubic centimeter. As the topic has matured, metrics using this technique have been developed to calculate the nanosheet thickness, length and concentration of semiconducting materials with an optical band-gap.[14,15] Unfortunately, the exact data analysis procedure used is not always trivial and erroneous values can result from over- or under-processing spectral data.[16] We have produced a simple computer program to reproducibly complete this analysis for common transition metal dichalcogenides (TMDs), and other semi-conductors with excitonic behavior to output average nanosheet dimensions.

In the first part of the manuscript, we provide a brief tutorial on the relevant theory for nanosheet characterization by optical spectroscopy. We shall then examine the challenges that must be overcome to accurately use UV/VIS extinction spectroscopy and discuss our solutions to address these. A software solution to implement these solutions is also described and made freely available on GitHub: *https://github.com/S-Goldie/A2DfromUV*.

### 1.1. Theoretical Review

When liquid dispersions of 2D materials were first reported, initially of graphene and then other layered materials including $WS_2$ and $MoS_2$, the optical extinction spectra were measured and attempts were made to establish extinction coefficients.[15,17] Since those reports, the study of nanomaterials has progressed and found different size dependent spectral features. One crucial fact to be highlighted is that common UV/VIS spectrometers measure the transmission of light through a sample to calculate the extinction (*Ext*) which contains contributions from both absorbance (*Abs*) and scattering (*Sca*) according to Equation 1:[18]

$$\log\frac{I_0}{I} = Ext = Sca + Abs \qquad (1)$$

Where $I_0$ is the intensity of the incoming light and $I$ the intensity after penetration through the sample.

In solutions of small molecules and ions scattering is negligible, so the distinction between true absorbance and the more commonly measured extinction is little more than semantic. However, for 2D nanosheets with sizes approaching or exceeding the wavelength of light, such scattering is far from negligible.[18] Fortunately, scattering spectra follow absorbance in shape (albeit red-shifted) so all key transitions are present in extinction. Nevertheless, the spectral profile and contributions are distinct between them.[15]

Optical transitions measured in this way have proven useful because of the systematic changes that occur as nanosheet size changes. These are best visualized in a set of samples with different nanosheet sizes, for example produced by liquid cascade centrifugation,[19] and are illustrated in **Figure 1** using size-selected $MoS_2$ as an example.

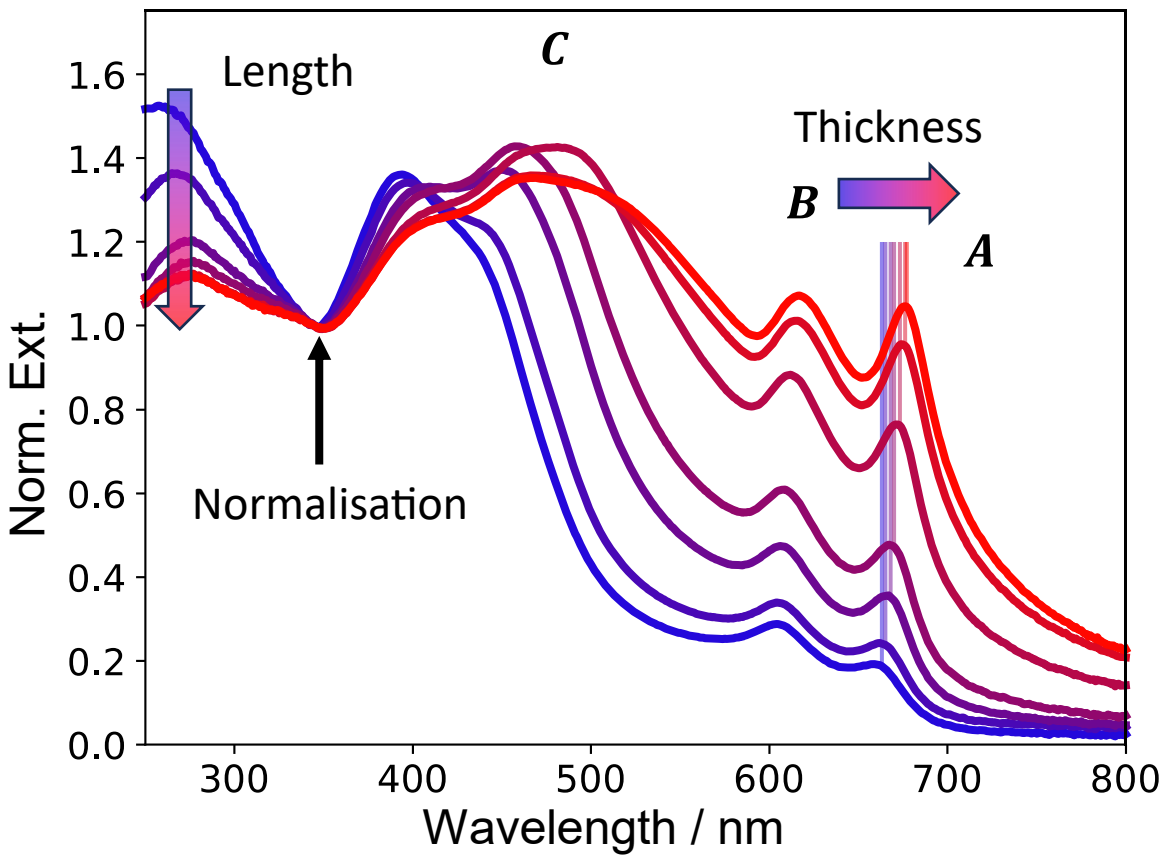

**Figure 1:** Normalized extinction spectra of liquid cascade size-selected $MoS_2$ nanosheet dispersions. From red to blue, the lines reflect fractions extracted at higher centrifugation rates leading to the smallest flake size shown in blue.

Nanosheet thickness can be determined from the position of distinct optical transitions.[14] The lowest energy of these, the A-exciton, corresponds to the optical band-gap, $E_A$, and has a significant layer number dependence because of the excitonic nature of the transitions in TMDs. The same trends discussed here are observed from the B and C excitons, and for excitonic transitions in other semiconducting materials which allows us to relate nanosheet thickness to the spectral position, *i.e.* energy of the excitons.[20–22]

Although nanosheet thickness is usually more useful, length metrics can also be derived from changes in intensity ratios at a set of wavelengths.[15] These are related to the difference between flake edges and basal plane center; which are assumed to have different absorption profiles. Thus, the relative extinction values at different wavelengths can be used to estimate average lateral sizes.

#### *1.1.1. Excitons and Nanosheet Thickness*

Understanding the thickness trends in 2D semiconductors requires a consideration of band theory and photoexcited states, specifically excitons. When a photon is absorbed, an electron is excited into the conduction band leaving an unoccupied electron state in the valence band. To simplify this many-body problem, rather than considering the large number of electrons remaining in the valence band, it is easier to consider a single positively charged state left behind, typically considered as a quasi-particle denoted a hole.

A simple interpretation of band-theory tells us the energy of this transition is equal to the optical band gap. However, in many semiconductors, these two oppositely charged quasi-particles are attracted to each other, and in a semi-classical picture can be considered analogously to a hydrogen atom with the excited electron bound to the positive hole state.[23]

This bound system is termed an exciton and is characterized by a stabilizing reduction in energy when compared to 'free' excited particles without any favorable interaction, this stabilization is termed the exciton binding energy ($E_b$). Employing the hydrogen atom model, the stabilisation energy can be considered equivalent to the ionization energy required to excite an electron from its orbital level. Within the band structure model this exciton binding energy reduces the energy of the system from the 'free particle' energy level, illustrated in **Figure 2**. This results in a lower optical transition energy ($E_A$) to excite an electron into a bound exciton state, as compared to the 'free particle' band gap ($E_g$). This stabilisation is given by:

$$E_A = E_g - E_b \quad (2)$$

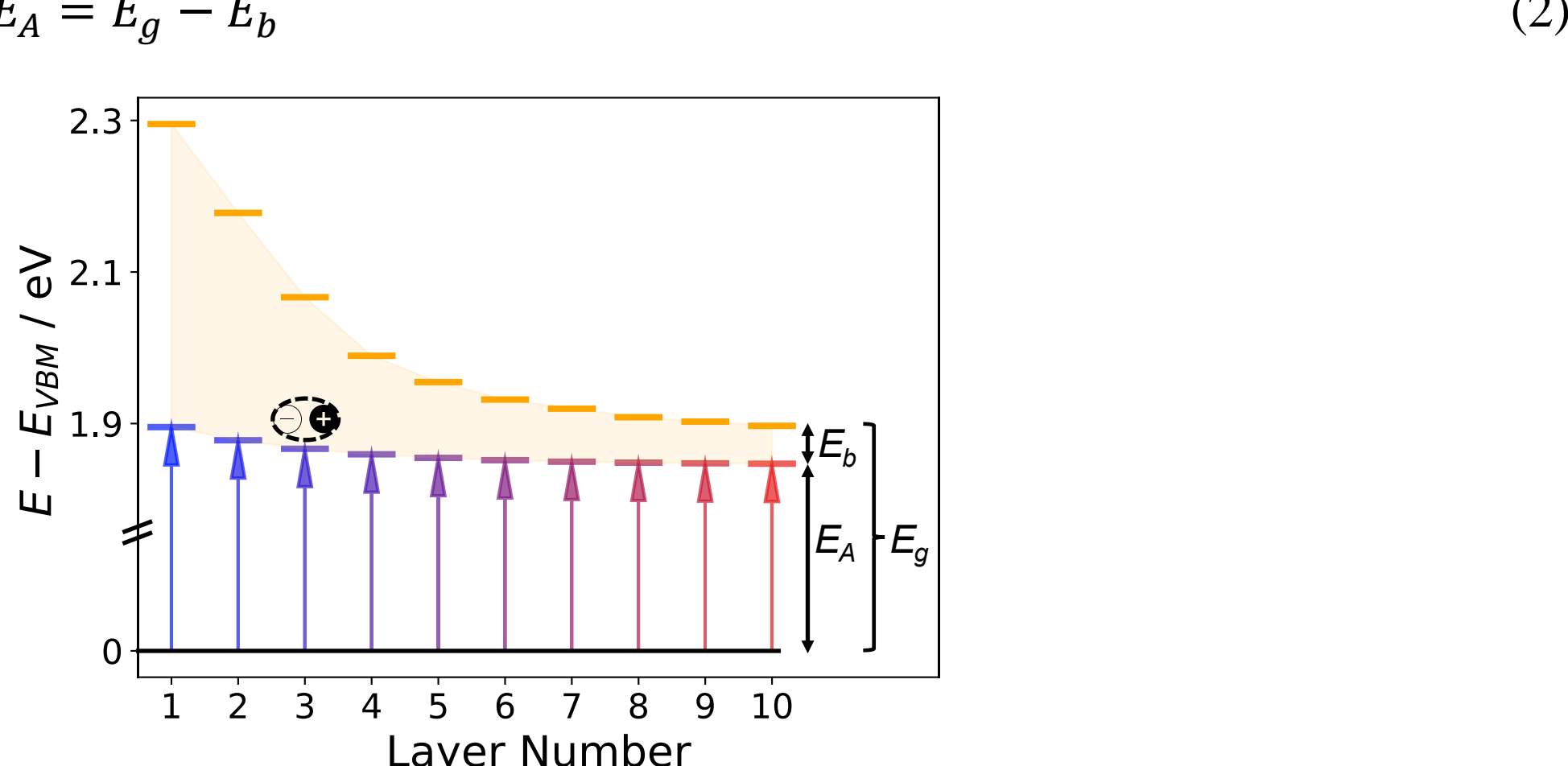

**Figure 2:** Schematic of the shift of the optical band-gap, $E_A$, typical in TMDs as layer number increases. Energy levels are not exact due to different values reported in literature, however the energy range shown is representative of the shifts found from $MoS_2$.

To understand the spectral response, the important question is how the optical transition energy is affected by the nanosheet thickness. The dominant influence on the energy levels is the long-range direct Coulomb interaction.[24] Considering the exciton binding energy ($E_b$), this attraction between charged particles depends on the dimensionality of the system and the dielectric environment in which the wavefunctions extend. As the system dimensions are reduced, exfoliating layers going from bulk 3D crystals into 2D nanosheets, the separation between charged particles is reduced, so the attractive interactions are felt more strongly and $E_b$ increases. This phenomenon is known as the quantum confinement effect.

Additionally, as layers are removed and replaced by media with different permittivity values, the dielectric screening between the charged particles is altered.[25] Permittivity is a materials' ability to rearrange its charge distribution in response to an electric field. Considering charged particles, they will attract a local charge density of opposite sign to themselves which acts to reduce the strength of the field experienced by other charged particles, outside the localized shell. This reduction in effective field strength is known as dielectric screening. If the screening decreases, for example by exfoliating layers from a bulk crystal so they are instead surrounded by air, the attractive force will increase and $E_b$ will again increase.

While the attractive Coulomb interactions cause an increase to the binding energy as layer number decreases; there is also a renormalization of the free-particle band gap ($E_g$). For these 'free' electrons, unbound to holes in the valence band, the long-range direct Coulomb interaction can be considered similarly to a classical repulsion of like charges. Thus, the quantum confinement effect and change in dielectric screening cause an increase to $E_g$ with decreasing layer number as the unfavorable repulsion is felt more strongly. Despite the large absolute magnitude of these changes, the net result on the optical bandgap ($E_A$) is remarkably small,[26,27] illustrated in Figure 2. Nevertheless, absorption and photoluminescence spectra have confirmed a regular change in optical band gap with layer number for semi-conducting 2D materials.[28,29]

Using liquid-exfoliated samples and a size selection process, an exponential decay in optical bandgap was found when going from bulk ($E_{A,bulk}$) to monolayers ($E_{A,ML}$):[14]

$$E_A = E_{A,bulk} + \left(E_{A,ML} - E_{A,bulk}\right)e^{(\langle N\rangle_{vf}-1)/N_0} \qquad (3)$$

For the optical band gaps $E_A$, $N_0$ is an empirical decay constant for the decrease in energy from monolayer to bulk and $\langle N\rangle_{vf}$ is the volume fraction weighted averaged layer number:

$$\langle N\rangle_{vf} = \frac{\sum N^2LW}{\sum NLW} \qquad (4)$$

where $N$ is the number of layers of a nanosheet, $L$ is the length and $W$ the width.

Equation 3 can estimate the average number of layers for a liquid-exfoliated sample as measured by absorption, *i.e.* excluding scattering. In extinction the overlapping contribution from scattering causes an additional shift in apparent peak position. We therefore find a more robust analysis can be completed using a slightly modified expression, discussed in Results and Discussion.

To conclude our discussion of the optical band gap for flake thickness determination; the uncertain influence of solvent permittivity must be acknowledged. By exfoliating layered materials and suspending them in vacuum, with a lower relative permittivity than the bulk material, the dielectric screening is reduced and both $E_b$ and $E_g$ are increased.[24,30,31] However, many solvents have relative permittivity values greater than the bulk material so the dielectric screening would be expected to increase rather than decrease. Calculating the exact magnitude of this shift has proven difficult and there remain conflicting accounts of the influence different solvents have.[14,15,32,33] Generally consistent thickness trends have been observed, but a cautious application should restrict metric use to solvents with similar permittivity values.

Permittivity notwithstanding, the change in A-exciton energy is currently understood according to a balance between quantum confinement and greater environmental influence on thinner sheets; and can be used to measure average nanosheet thickness of an ensemble in liquid dispersion. We note that similar trends have also been observed for substrate-supported mechanically-exfoliated sheets.[21,34,35]

*1.1.2. Length Metrics*

Another measurement of 2D nanomaterials is lateral nanosheet size. When considering liquid phase-exfoliated materials, as a first approximation length-to-width and length-to-thickness aspect ratios can be treated as material dependent properties. This is because the exfoliation process is controlled by the balance of inter- and intra-layer bond strengths.[13] As such, knowledge of layer number allows one to indirectly infer average lateral size for a given material exfoliated by sonication-assisted LPE.

For more sample specific analysis however, the change in optical spectra with lateral size has also been investigated. Hypothesizing every flake has an edge region, of thickness $x$, with a different electronic environment, density of states and therefore absorption coefficient, Coleman *et al.* suggested forming a new area-weighted average absorption coefficient:[15]

$$\alpha \approx \alpha_C + \frac{2x}{L}\left(\frac{L}{W} + 1\right)(\alpha_E - \alpha_C) \quad (6)$$

using the long dimension ($L$) and aspect ratio ($\frac{L}{W}$) of arbitrarily shaped 2D flakes and absorption coefficients at the edge ($\alpha_E$) and center ($\alpha_C$).

Assuming an effective edge region of 12 ± 7 nm (as experimentally suggested in [15]), this expression adequately describes nanosheets larger than 90 nm. Below this, edge effects become so dominant that the area-weighting fails to describe the system. While descriptive, to measure the flake length this relationship would require absorption coefficients at edge and center to be calculated- rarely a trivial task. Therefore, a more user-friendly formulation can be considered using the ratio of absorbance, or extinction values at different wavelengths and empirical constants $A_1$, $A_2$ and $B_1$:

$$\frac{ext(\lambda_1)}{ext(\lambda_2)} = \frac{A_1\langle L\rangle + B_1}{A_2\langle L\rangle + 1} \quad (7)$$

The empirical constants have been measured and reported previously for all of the semiconducting nanomaterials included in our library, and the constants are listed in Supporting Information S1.

Equation 7 fits well for flakes between $70 < \langle L \rangle < 350$ nm, ideal for samples prepared by liquid phase exfoliation although new methods like electrochemical exfoliation may produce larger nanosheets. This expression and constants are included within our analysis program for convenience, but no refinement is necessary.

*1.1.3 Scattering*

Both length and thickness were originally investigated using absorbance data because of the clearer discussion without the scattering background. Nevertheless, for many dispersions the key transitions are found in both extinction and absorbance spectroscopy so with suitable corrections, extinction spectra can be used.

In the non-resonant regime, that is at lower energy below any resonant optical transitions, the scattering ($\sigma$) has been investigated and shown to follow a power law dependency to wavelength ($\lambda$):[18]

$$\sigma(\lambda) \propto \lambda^{-m} \qquad (8)$$

This scattering can be described by a generalized equation for a transition between Rayleigh scattering ($m = 4$) that applies to smaller nanosheets and the van de Hulst approximation of the more general Mie scattering theory ($m = 2$) that applies for larger nanosheets. This trend is illustrated in **Figure 3**, flake lengths were estimated from Equation 7 and the different exponents ($m$) fitted in the linear region below 850 nm follow the trend between mostly Rayleigh scattering at 80 nm towards Mie scattering at 294 nm.

This scattering can be described by a generalized equation for a transition between Rayleigh scattering ($m = 4$) that applies to smaller nanosheets and the van de Hulst approximation of the more general Mie scattering theory ($m = 2$) that applies for larger nanosheets. This trend is illustrated in **Figure 3**, showing extinction spectra of $MoS_2$ nanosheets on a double logarithmic scale to visualize the power law scaling. Average flake lengths as indicated in the figure were estimated from Equation 7. The different exponents ($m$) fitted below 850 nm follow the trend between mostly Rayleigh scattering for nanosheets with an average length of $\langle L \rangle$ = 80 nm towards Mie scattering at $\langle L \rangle$ = 294 nm.

This size-dependent exponent can estimate platelet volume if sufficient data can be recorded between the lowest energy optical transitions and the near-IR vibrational modes of the solvent. More importantly for our discussion, however, power law behavior only applies away from the optical transitions on which we base our thickness and length metrics. The failure of extrapolating the power law is shown in Figure 3, illustrating the large deviation encountered for the samples containing larger nanosheets on average.

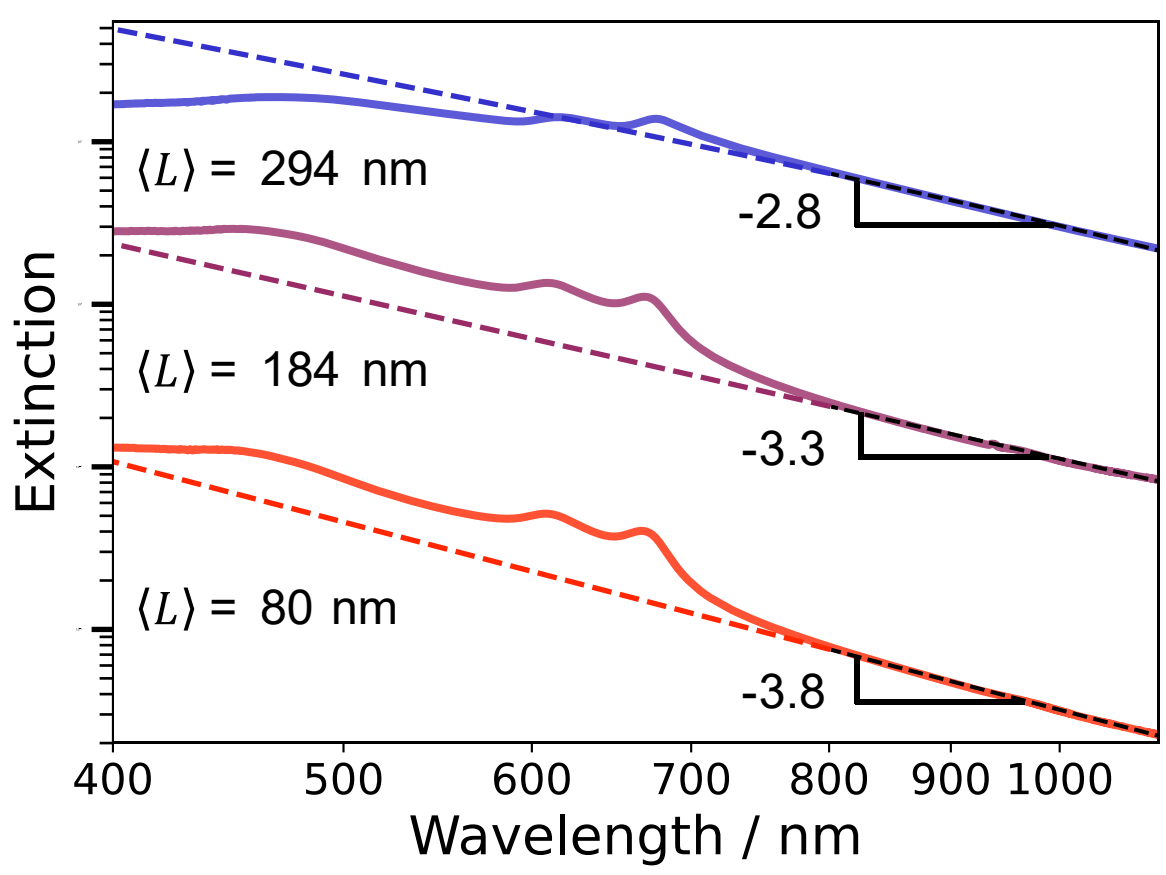

**Figure 3:** Extinction spectra of $MoS_2$ on a double logarithmic scale. Power-law fits in the non-resonant region are shown as dashed lines. Each spectrum is shown offset for clarity. The samples decrease in nanosheet size (length measured using Eqn.7) from blue to orange which results in a subtle change in the exponent of the fit ($m$), labelled next to each spectrum.

Finally, the concentration of nanosheet dispersions is a key parameter for many processing steps. As discussed, the absorbance profile depends on nanosheet size but for some materials specific wavelengths have been identified where the size effects balance and the absorption remains relatively invariant to flake size. Using the absorption, or extinction, at these wavelengths, coefficients have been identified that can be used to calculate nanosheet concentration according to the Beer-Lambert Law:

$$Ext = \varepsilon c l \qquad (9)$$

Extinction coefficients, and the wavelengths at which they are valid, are included in **Table 1** and for convenience, concentrations are automatically calculated within our algorithm. The wavelength at which extinction is relatively invariant to flake size is also used as the normalization point for plotting, since this value illustrates the difference between larger and thinner nanosheets very clearly, as seen in Figure 1.

With the theoretical background and known relationships outlined, we now turn to optimizing the analytical process and present an automated program capable of reproducibly completing such analysis.

**Table 1**: Extinction coefficients and wavelength used to estimate the concentration of nanosheets.

| Material | $\varepsilon$ [L g$^{-1}$ cm$^{-1}$] | $\lambda$ [nm] |
|---|---|---|
| $WS_2$ [19] | 48 | 235 |
| $WSe_2$ [14] | 40 | 440 |
| $MoS_2$ [15] | 69 | 345 |
| $MoSe_2$ [14] | 50 | 358 |
| $RuCl_3$ | 14.8 | 500 |
| $NiPS_3$ [36] | 12.5 | 383 |
| $PtSe_2$ | No size independent wavelength found | |
| γ-InSe [37] | 53 | 341 |

## 2. Results and Discussion

To make use of these metrics, extinction values at specific wavelengths must be read and accurate wavelengths of the exciton transitions must be found. For simple molecular systems this can often be done by finding the extinction maxima. A more involved approach would fit peak models to the spectrum. Unfortunately, these are unreliable when applied to extinction spectra of nanosheets because of the uncertain scattering background and overlapping contributions. While good fits can be obtained, caution must be employed.

In the following section we will outline the key reasons for these difficulties and explain the solutions proposed to solve this analysis problem in an automated way thus making the analysis independent of the user. The new approach we take is to apply a smoothing algorithm to each spectrum to reduce the effects of noise. Once a suitable smoothing value has been identified, the second derivative is computed and the peak center of the A-exciton transition found. By applying this method to previously published spectral and microscopy datasets, we have refined thickness metrics for semi-conducting materials.

As discussed in the introduction, there is no reliable model for scattering in the resonant regime, so an approximation like a polynomial or local power-law background function would have to be used. In many materials there are also overlapping peaks. $MoS_2$ shows this very significantly due to the small difference in energy (0.15 eV) between the A- and B-excitons making it a useful test case.

Fitting an $MoS_2$ spectrum can produce a visually pleasing fit, example illustration in **Figure 4a**. However, the non-linear fitting is very susceptible to peak overlap. This overlap can be roughly quantified by the correlation matrix between fitted parameters. When directly fitting the spectrum, the A-exciton position has a correlation coefficient of 0.41 with the B-exciton position, and 0.46 with the B-exciton peak width. This means that any uncertainty or error experienced with the B peak fitting will alter the apparent A-exciton energy. This is not to say that measurement and peak fitting in this way cannot be carefully undertaken by a skilled user, but such difficulties preclude this method as a universally applicable method for material characterization.

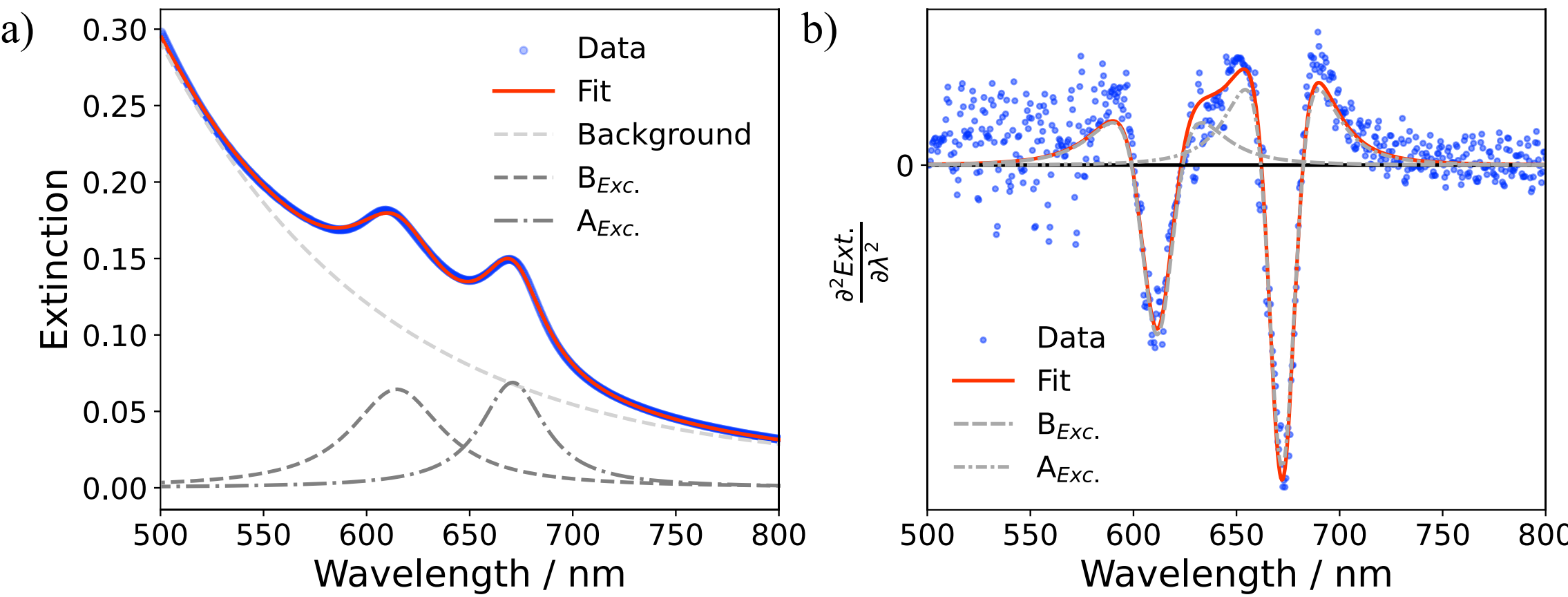

**Figure 4:** Example fitting of A and B excitons of $MoS_2$ in a typical extinction spectrum (a) and following differentiation (b). Before differentiation the two peaks, visualized by the grey dashed traces, are seen to overlap and correspond less to the intensity of the spectrum than the approximate background model. After differentiation the peaks are better separated: the only peak overlap is in the positive region and the peak intensities of the grey dashed traces dominate over the now negligible background.

One solution to sharpen peaks is to use the second derivative. This leads to better peak separation and numerically equates to a second order polynomial subtraction.[38] Even visually, the improvement is clear in Figure 4b: the two peaks are now well separated, with only slight overlap of the tails in positive differential space. Further, the y-range is dominated by the peak intensity rather than background trends. Fitting the same model in $\partial^2/\partial\lambda^2$ form shows the correlations of A-exciton position to other fit parameters is reduced to < 0.05 and the estimated error is reduced by 50%. Therefore, taking the derivative resolves many of the issues encountered when analyzing the spectrum directly.

Unfortunately, since numerical derivatives are highly sensitive to noise, smoothing is necessary. We have found that locally weighted scatterplot smoothing (LOWESS) is well suited to these data sets. This smoothing method uses linear regression to locally fit data and estimates smooth values. The local function is fitted over a number of points that are described as the 'smoothing window'. As the smoothing window increases more data points are included and the greater the effect of the smoothing function. Smoothing is effective at reducing noise in the second derivative plot, but the sloped scattering background and superposition of other excitons, where present, can lead to an apparent shift of peak intensity with stronger smoothing. Depending on experimental conditions and data acquisition, different spectra require different degrees of smoothing, but if incorrectly applied, such smoothing can also cause error in the analysis. Fortunately, attempts to quantify this effect also provided a possible solution to the issue.

Considering the apparent A-exciton wavelength as smoothing was changed, most spectra show instability for insufficient smoothing. This is visualized in **Figure 5a** using the same sample measured at different concentrations. Despite the identical material and acquisition parameters, a certain minimal level of smoothing is required for consistent analysis, which is different for each concentration as indicated by the colored dashed lines in Figure 5a. Once this minimum smoothing window is exceeded, a relatively consistent A-exciton position is found as the smoothing window increases, before even greater smoothing causes a regular erroneous shift to the peak position, away from the consistent value highlighted by the horizonal dashed line. We therefore endeavor to find this stable region, applying just enough smoothing to allow reproducible analysis. The importance of this is illustrated by the analysis of the lowest concentration dispersion, which converges to the same A-exciton position as the more concentrated samples before diverging to higher apparent wavelengths as greater smoothing is applied.

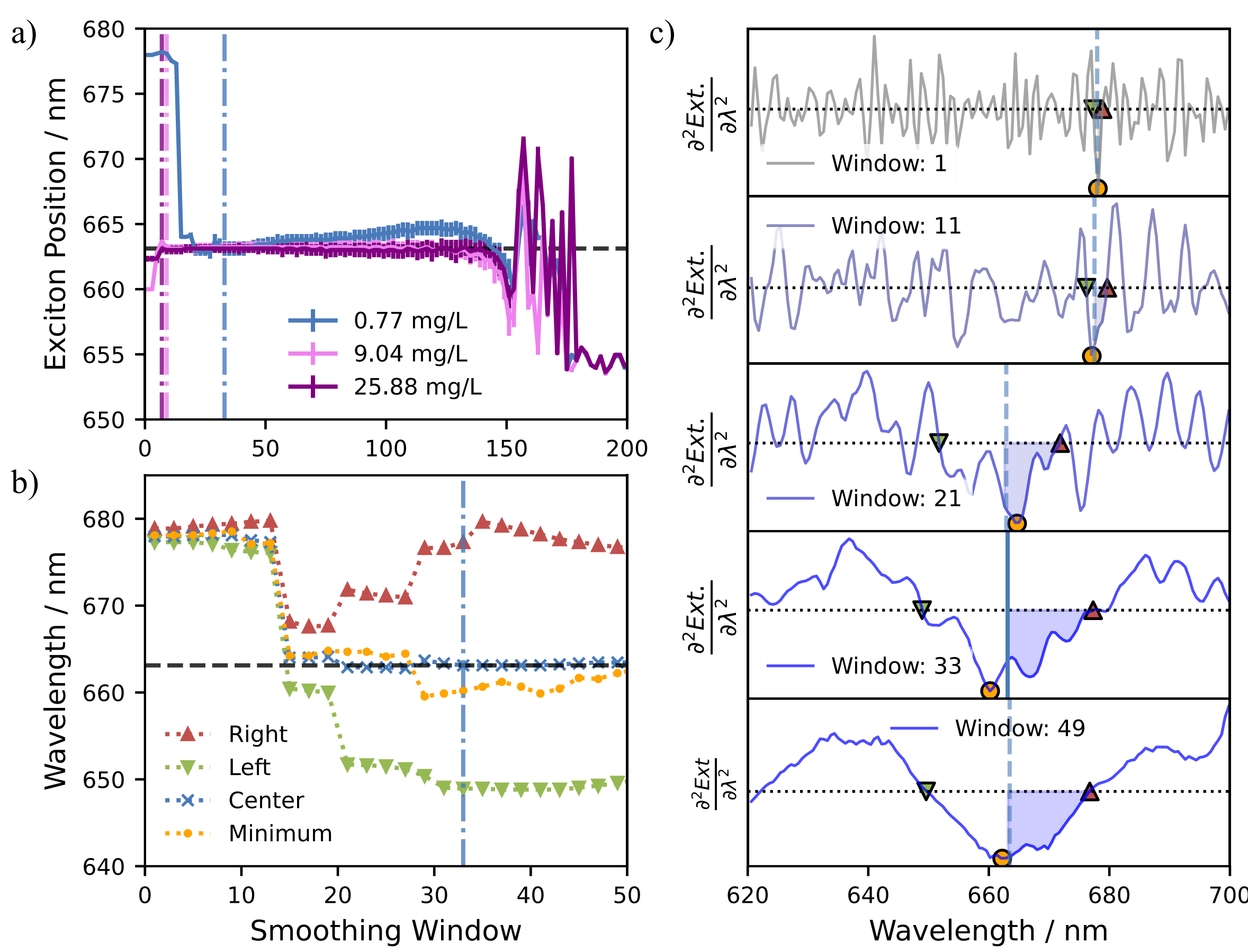

**Figure 5:** Key requirements for the algorithmic process for peak analysis. a) The measured exciton transition wavelengths that result from different smoothing windows for an identical sample recorded at differing dilutions. The consistent wavelength common to all dilutions is marked by the horizonal dashed line. Vertical dashed lines show the minimal smoothing required according to our automated analysis process. b) Variation of key spectral metrics for the most dilute 0.77 mg/L spectrum: right (▲) and left (▼) intercepts with the x-axis, minimum between these intercepts (●) and peak center (x). Dashed lines denote the same as in (a): horizonal is the consistent transition wavelength and the vertical the minimal smoothing according to our algorithm. c) The effect of increasing smoothing window on second derivative spectra and the intercepts and peak minimum used to identify the suitable analysis regime. Integral areas are highlighted by the shaded areas under the curves.

To find this region, we utilized an algorithm sequentially increasing smoothing and checking each smoothed, differentiated spectra for stability as described below. For each smoothed, double differentiated spectrum, the $\partial^2/\partial\lambda^2$ minimum, and from there the closest x-intercepts, are found. The respective values are plotted in Figure 5b as a function of smoothing window. If the intercepts are less than 15 nm apart from each other, or within 5 nm of the peak minimum, random noise is still dominating as seen from the lowest smoothing windows in Figure 5c.

In each smoothed second derivative in Figure 5c, minimum and closest x-intercepts are indicated by the colored symbols. For higher smoothing windows, e.g. the spectrum with 33 points per window, a visually apparent peak begins to form, albeit with some noise that can still strongly affect the minimum. This effect becomes negligible at higher smoothing levels, e.g. 49 points per window. However, since a gradual deviation from the stable peak position was observed in Figure 5a beyond 50 points per window, it is important to apply as little smoothing as possible.

The overall trends of the minimum and x-intercepts are shown in Figure 5b. For very small smoothing windows the initial tight bunching is seen before the true peak limits are established and the minimum and intercepts begin to follow a more regular trend with changing smoothing window. From our experience, attempting this method with 100+ spectra, we identify a stable smoothing window when changing the smoothing window causes less than 2.5 % variation to the intercepts and peak minimum. The criterion of 2.5 % is somewhat arbitrary. However, significantly less than this value results in very large smoothing windows that destroy the true peak shape. When using greater tolerance than this, we found the x-intercepts became unstable for asymmetric peaks, like monolayer $WS_2$, or broad peaks as recorded for nanosheets larger than 200 nm. For a more detailed discussion on these criteria with data sets and plots, see Supporting Information S3.

Note it is possible that dilute samples, with maximum extinction values <~0.05, require longer experimental dwell times to improve signal to noise for this stable analysis window to be found. However, we also note that extinction values >> 0.05 at the excitonic transitions are typically accessible for LPE nanosheets.

Provided the minimal sufficient smoothing has been determined, the only problem remaining is to determine the position of the exciton transition correctly. For dispersions with relatively narrow thickness distributions in the few-layer regime, peak fitting can be applied. However, this is often not the case. In many real samples asymmetric peak broadening or even peak splitting of the A-exciton as in $WS_2$ is observed. The splitting of the A-exciton in $WS_2$ is a special case and has been reported previously as an additional metric of monolayer content;[16,19] for a more comprehensive discussion see Supporting Information S2. When these split or asymmetric peaks are fitted separately, the difficulties in deconvoluting two or more overlapping peaks would become even more pronounced than in the example spectrum of $MoS_2$ discussed above.

To overcome the issue with irregular peak shapes and experimental backgrounds, we suggest a simple method for peak position determination that is insensitive to peak shape. Rather than using the minimum in the second derivative, the peak center of mass is determined from numerical integrals in the second derivative spectrum and the x-axis intersections are used as left and right limits. The area enclosed between the curve and the axis is then determined and the center of mass position is the wavelength which divides the area into two equal parts, as illustrated in the bottom panels of Figure 5c as blue shaded area and vertical blue line. This center of mass position is also included as a function of smoothing window in Figure 5b and is much more robust in terms of peak position even for spectra below the minimal smoothing window, where a peak in the second derivative is hardly deciphered by the human eye, e.g. 21 points per window.

The overall sequence of the algorithm-based peak finding is outlined in **Scheme 1**. As outlined in the theoretical review above, the peak position can be related to the average layer number using previous correlations to the nanosheet thickness measured by AFM that serve to establish a calibration curve for each material.

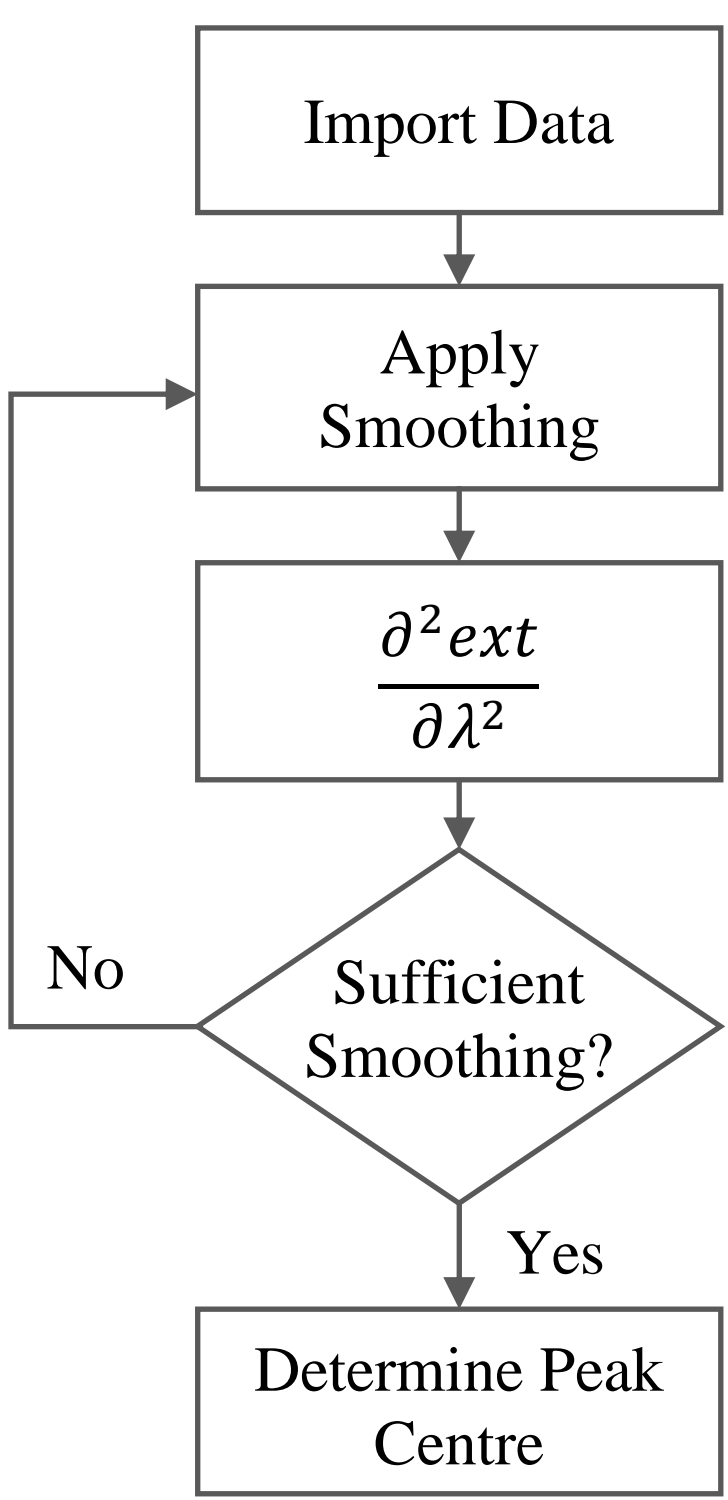

**Scheme 1:** Flowchart illustrating the key processes in the algorithmic analysis of UV/VIS spectra; the criteria for sufficient smoothing is a < 2.5 % variation in peak parameters as the smoothing window is changed.

Previous reports have measured metrics of flake thickness, often with a focus on absorbance spectra that are acquired in the center of integrating sphere where scattered light is collected. This typically resulted in a logarithmic function to calculate the layer number from the exciton wavelength with a sharp limit of "bulk" behavior around 10 layers.[14,39] While valid for absorbance, the additional influence from scattering in extinction spectra often means larger flakes exhibit a greater red-shift in apparent flake position that correlates to larger, thicker flakes.

Since extinction measurements are more practical than absorbance, we present empirical correlations of flake thickness to exciton energy in extinction below. To this end, we gathered all our available data of extinction spectra of samples where nanosheet sizes were determined by AFM statistics, applied the algorithm-based analysis described above and plotted $\langle N \rangle_{vf}$ as a function of exciton energy to find the most suitable empirical function. This is shown for $MoS_2$ in **Figure 6a** and $WS_2$ in Figure 6b. For other materials, see Supporting Information S4.

We find excellent agreement with an exponential function of the form

$$\langle N \rangle_{vf} = N_{bulk} \cdot e^{R(E_A - E_{A,bulk})} \quad (10)$$

Here, $E_{A,bulk}$ are energies reported in literature (see **Table 2**). We then fit an 'apparent' layer number ($N_{bulk}$) and decay constant ($R$) for each material. For further discussion of the form of this expression see Supporting Information S3.

This empirical expression has been successfully applied to a library of eight 2D semi-conductors where spectra and nanosheet size statistics, from AFM studies, have been published previously.[14–16,36,37,40–43] The fitted constants that result are given in Table 2. The fits and the error of the fit are shown in Figure 6a, b and for other materials in Supporting Information S4.

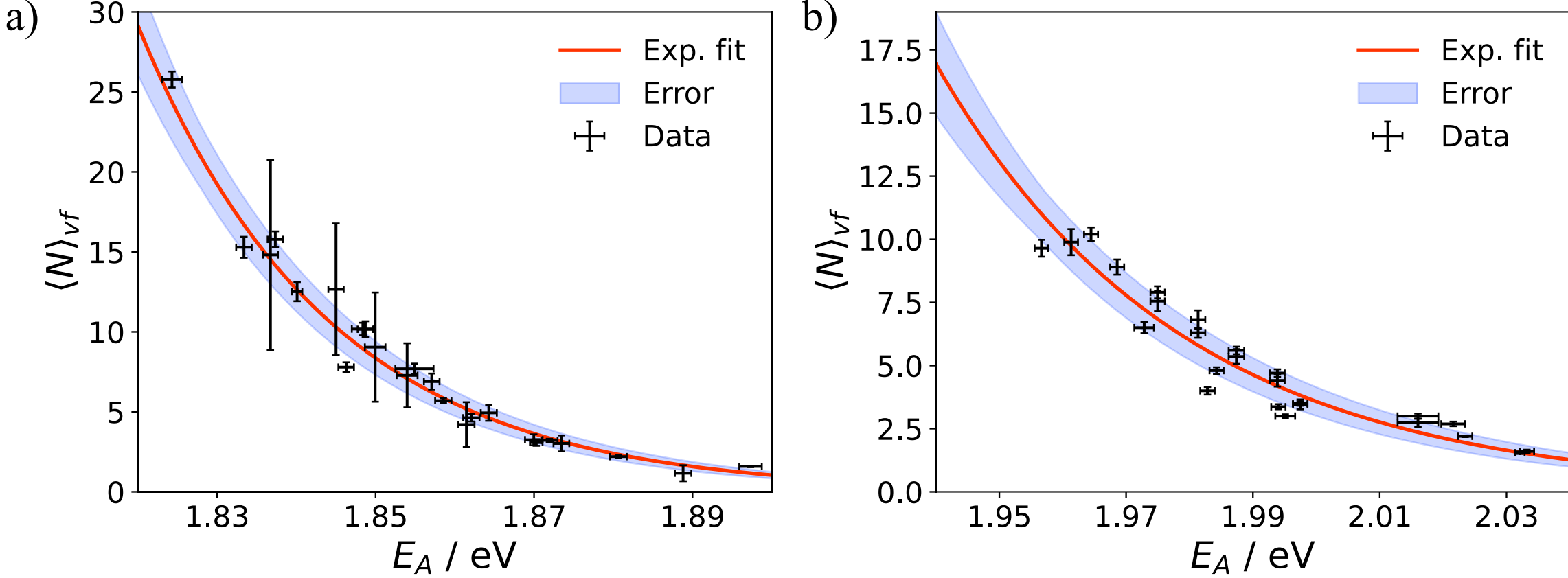

**Figure 6:** Plots of $\langle N \rangle_{vf}$ as a function of exciton energy found from the algorithm-based analysis from $MoS_2$ (a) and $WS_2$ (b). All data used in this plot was gathered in our laboratory and published previously.[14–16,40] Errors in $E_A$ are from the analysis process introduced in this publication, errors in $\langle N \rangle_{vf}$ are statistical standard errors from the AFM measurement. An exponential decay, shown in red with the fit confidence area shown in pale blue, describes the datasets well.

This exponential decay was fitted with an orthogonal fitting procedure to weight points according to errors in both measured exciton transitions and statistical errors from AFM studies.[44] Confidence intervals of the metric are shown as a blue shaded area and uncertainties of the fitted constants are included in Table 2.

By propagation of errors these uncertainties can be combined with the uncertainty in measured exciton energy to provide a confidence interval of thickness determination. The error in spectral analysis is estimated from the convergence of the integral areas to a minimum, with a correction for signal to noise detailed more in Supporting Information S5.

**Table 2:** Fitted constants for a library of materials. $WS_2$ and $MoS_2$ are more widely published providing a greater data set for fitting. The fitted trends for $MoS_2$ and $WS_2$ are shown in Figure 6, fitted trends for all materials are available in Supporting Information S4.

| Material | $N_{bulk}$ | $R$ | $E_{bulk}$ |
|---|---|---|---|
| $MoS_2$ | 20.9 ± 1.0 | - 41.6 ± 1.5 | 1.828 [14] |
| $WS_2$ | 10.9 ± 0.7 | - 25.9 ± 1.5 | 1.957 [14] |
| $MoSe_2$ | 26.3 ± 3.6 | - 41.9 ± 3.7 | 1.516 [14] |
| $WSe_2$ | 14.2 ± 2.7 | - 28.9 ± 4.7 | 1.610 [14] |
| $PtSe_2$ | 34.1 ± 5.0 | - 12.3 ± 0.9 | 1.739 [a] |
| $RuCl_3$ | 11.5 ± 0.7 | - 24.3 ± 2.0 | 3.235 [37] |
| $NiPS_3$ | 36.0 ± 3.3 | - 9.6 ±1.0 | 3.139 [37] |
| γ-InSe | 28.4 ± 0.7 | - 6.0 ± 0.1 | 3.101 [a] |

a) The analysis described in the work used different transitions than those used in previous metrics.[37,43] To calculate a suitable $E_{bulk}$ value an identical analysis to the literature reports was used for the different transitions.

## 3. Conclusion

We have developed an algorithm for the analysis of optical extinction spectra of semiconducting nanomaterial dispersions and implemented it within a computer program to output nanosheet dimensions. While the determination of average lateral size from peak intensity ratios in the extinction spectra is straight-forward, an accurate analysis of peak positions, which relate to average nanosheet thickness, can be challenging.

To address this, we have implemented the following innovations: by using the axis intercepts and minimum value in second derivative space we can identify the minimum signal smoothing required for consistent analysis. The spectrum thus smoothed can be analyzed in second derivative space by finding the peak center of mass, in place of minimum values, thereby avoiding potential errors caused by fitting inappropriate models to more complex peak shapes. This analysis of relevant spectral features, usually the A-exciton, was then used to refine metrics to allow rapid determination of average flake thickness from UV/VIS extinction spectra which are more accessible than absorbance spectra measured with an integrating sphere that were often used in literature before.

This algorithm is made freely available for download and can be applied to a range of semi-conducting materials. By sharing the methodology to develop the thickness metric from peak positions extracted in this manner, we also invite others to add new materials to this library. We hope that such a procedure can then be applied to rapidly screen dispersions of 2D materials for required nanosheet dimensions reliably and consistently prior to additional use.

**Methods**

Metrics were fitted to literature data published previously where UV/VIS spectral data and full AFM counts were available. All fitting and data analysis was completed with Python3. To fully account for the statistical error in AFM counts and the fitting error in spectral analysis an orthogonal fitting procedure (`scipy.odr`) was used minimizing the geometric distance between the line of best fit and available data points, rather than more conventional linear regression which only accounts for difference in the y-direction.[44] All errors from fitted parameters are standard deviations as defined from model fitting. Statistical errors from AFM measurements are standard errors.

Where example spectral data is shown, this was collected from freshly exfoliated $MoS_2$ stabilized in aqueous sodium cholate prepared by sonication-assisted liquid phase exfoliation and cascade centrifugation to isolate large/thick flakes as sediment:[19]

*Materials*

Molybdenum (IV) sulfide power (<2 μm 99%), tungsten (IV) sulfide power (< 2 μm 99%), and sodium cholate hydrate from bovine and/or ovine bile ≥99% were purchased from Merck Sigma-Aldrich and used as provided. Molybdenum (IV) selenide (325 mesh powder, 99.9%) was purchased from Alfa Aesar and used as provided.

*Exfoliation*

TMD powder (20 mg/mL) was probe sonicated in an 80 mL aqueous solution of sodium cholate (8 mg/mL) for 1 hour of active sonication time (pulsing 6 s on and 4 s off) using a solid flathead tip (Sonics VXC-500) set to 60% amplitude, at a controlled external temperature of 5 $^{o}$C. During all sonication the probe was placed approximated 1.5 cm above the bottom of the metal beaker. The dispersion was centrifuged in 20 mL aliquots using 50 mL vials in a Hettich Mikro 220R centrifuge equipped with a fixed-angle rotor 1016 at 3,820 g for 2 h. The supernatant was removed and the sedimented powder sonicated for a further 5 hours of active sonication time (pulsing 4 s on, 4 s off) in 80 mL aqueous sodium cholate (2 mg/mL). From our experience, this two-step sonication procedure yields a higher concentration of exfoliated material and removes impurities.

*Cascade Centrifugation*

The cascade used in this case separated into fractions isolated between 100 x *g*, 400 x *g*, 1000 x *g*, 6000 x *g*, 10,000 x *g* and 30,000 x *g*, where *g* corresponds to relative gravitational force. All centrifugation was done using a Hettich Mikro 220R with temperature control active and set to a nominal temperature of 10 °C. Fractions isolated at 100 x *g* and 400 x *g* used 20 mL aliquots in 50 mL centrifuge vials with a 1016 fixed-angle rotor; fractions isolated at 1000 x *g*, 6000 x *g*, 10,000 x *g* used 2 mL aliquots in 2 mL vials in a 1195-A fixed-angle rotor, the fraction sedimented at 30,000 x *g* used 1.5 mL aliquots in 1.5 mL vials in a 1195-A fixed angle rotor.

*UV/VIS/nIR Spectroscopy*

UV/VIS/nIR spectra were recorded on an Agilent Cary 60 in quartz cuvettes with a 1 cm pathlength. During development, different acquisition parameters were trialed to better understand measurement effects on the analysis, however for all spectral data displayed in this manuscript 0.6 nm increments were used with a dwell time of 0.1 s per datapoint.

**Supporting Information**

Supporting Information is available from the Wiley Online Library or from the author.

**Acknowledgements**

The authors wish to thank Kevin Synnatschke for supplying original datasets for analysis and Tim Nowack for useful discussions on UV/VIS experiments and the limitations thereof.

**Funding**

We acknowledge funding from the European Union's Horizon Europe research and innovation program under grant agreement number 101135196.

**Author Contributions**

Nico Kubetschek did formal analysis: lead; methodology: supporting; software: equal; visualization: lead; writing – original draft: equal

Claudia Backes did data curation: equal; funding acquisition: lead; supervision: lead; writing – review & editing: supporting

Stuart Goldie did conceptualization: lead; data curation: equal; formal analysis: supporting; methodology: lead; software: equal; supervision: supporting; writing – original draft: equal; writing – review & editing: equal

**Conflict of Interest**

The authors declare no conflicts of interest.

**Data Availability Statement**

The data that supports the findings of this study are available in the supplementary material of this article. Complete spectral data sets and Python code used within are available freely for download on GitHub: https://github.com/S-Goldie/A2DfromUV.